\documentclass{article}
\usepackage[utf8]{inputenc}
\usepackage[numbers]{natbib}
\usepackage{graphicx}
\usepackage{amsfonts}
\usepackage{amsmath}
\usepackage[table]{xcolor}
\usepackage{diagbox}
\usepackage{caption}
\usepackage{multicol}
\usepackage{array}
\usepackage{wrapfig}
\usepackage{geometry}
\usepackage{abstract}
\usepackage{soul}
\usepackage[hyphens]{url}

\usepackage{float}
\usepackage{authblk}
\usepackage{hyperref}

\DeclareMathOperator{\MSE}{MSE}

\graphicspath{ {Imgs/} }

\title{Physics-informed Machine Learning to predict Sea Surface Temperature using consistent Koopman and convolutional autoencoders}

\title{Analyzing Koopman approaches to physics-informed machine learning for long-term sea-surface temperature forecasting}

\author[1,2]{Julian Rice\thanks{Corresponding author. Email: \href{mailto:jrice15@calpoly.edu}{jrice15@calpoly.edu}}\textsuperscript{,}}
\author[1]{Wenwei Xu}
\author[3]{Andrew August}

\affil[1]{Marine and Coastal Research, Pacific Northwest National Laboratory}
\affil[2]{California Polytechnic State University, San Luis Obispo}
\affil[3]{Computing and Analytics, Pacific Northwest National Laboratory}

\date{July 2020}

\begin{document}

\maketitle


\begin{abstract}

Accurately predicting sea-surface temperature weeks to months into the future is an important step toward long term weather forecasting. Standard atmosphere-ocean coupled numerical models provide accurate sea-surface forecasts on the scale of a few days to a few weeks, but many important weather systems require greater foresight. In this paper we propose machine-learning approaches sea-surface temperature forecasting that are accurate on the scale of dozens of weeks. Our approach is based in Koopman operator theory, a useful tool for dynamical systems modelling. With this approach, we predict sea surface temperature in the Gulf of Mexico up to 180 days into the future based on a present image of thermal conditions and three years of historical training data. We evaluate the combination of a basic Koopman method with a convolutional autoencoder, and a newly proposed ``consistent Koopman" method, in various permutations. We show that the Koopman approach consistently outperforms baselines, and we discuss the utility of our additional assumptions and methods in this sea-surface temperature domain.

\end{abstract}


\vspace{3em}

\begin{multicols}{2}

\section{Introduction}

\subsection{Background}
\subsubsection{Long-term sea-surface temperature forecasting}
The sea surface temperature in the Gulf of Mexico is of great environmental, economic, and social significance. It regulates the low oxygen “dead zone”\cite{rabalais2002gulf}, fuels intense tropical cyclones \cite{shay2000effects}, such as Katrina and Harvey, and modulates thunderstorms in the southern US \cite{edwards1996comparisons}. The Gulf of Mexico also has unique flow features such as the Loop Current, which forms when warm subtropical waters intrude into the Gulf through the Yucatan Channel and eventually push out through the Florida Straits. The Loop Current periodically sheds warm eddies, which propagate westward in the Gulf for 9-12 months \cite{shay2000effects}, causing distinct temperature dynamics. East of Florida, much of the tail end of the Loop Current forms the Gulf Stream, another unique and important flow feature that pushes warm water into the North Atlantic.

Current state-of-the-art SST predictions are from atmosphere-ocean coupled numerical models, which do not make predictions beyond a month \cite{counillon2009high} \cite{oey2005exercise}: for example NOAA's Real Time Ocean Forecast System (RTOFS) only forecasts up to five days \cite{spindler2006noaa}. Additionally, numerical models are computationally expensive and rely on initial boundary conditions which are generally hard to obtain. It is also notable that the eddy shedding events which involve the rapid growth of nonlinear instability is particularly difficult to forecast in numerical models \cite{oey2005exercise}. 

In this study, we evaluate machine-learning (ML) techniques for long-range SST forecasting. More specifically, we well predict SST in the Gulf of Mexico up to 180 days in the future from a single thermal image of SST, leveraging techniques based in Koopman operator theory.

\subsubsection{Nonlinear dynamical systems}
For discrete-time dynamics problems, we seek a transformation $\varphi$ that maps a system's state $z$ at time $t$ its state at the next time step:
    \begin{equation}
    \label{eqn:dyn-sys-map}
        z_{t+1} = \varphi(z_{t})
    \end{equation}
where $S \subset \mathbb{R}^n$ is the space of all possible states, $z \in S$, and $\varphi: S \rightarrow S$. We focus specifically on systems where $\varphi$ is time-invariant; that is, $\forall t \in \mathbb{Z}$ equation (\ref{eqn:dyn-sys-map}) holds. 

In dynamical systems problems where $\varphi$ is nonlinear, a machine-learning approach is often appealing, since the system can be too complex to model analytically or numerically. However, a purely data-driven approach eschews physics knowledge that can be extremely helpful in developing meaningful and generalizable approximations of $\varphi$. Thus, we choose to utilize physics-informed machine-learning, whereby our understanding of physical systems is incorporated into machine-learning models. 

Additionally, we often do not have access to the state itself, only ``observables" of it, which are simply functions of the state that are much more easily measured:
\begin{equation}
    \label{eqn:observable}
    y_t = f(z_t)
\end{equation}
where $f: S \rightarrow \mathbb{R}$, and $y_t \in \mathbb{R}$ is the value of an observable $f$ of the system's state at time $t$. Thus, we need a more complex formulation than Eqn. (\ref{eqn:dyn-sys-map}).

\subsubsection{A Hitchhiker's Guide to Koopman Operator Theory}
Koopman operator theory \cite{koopman1931hamiltonian} has recently been \mbox{(re-)}discovered \cite{budisic2012applied} to be a valuable tool in modelling dynamical systems: it suggests that such a nonlinear dynamical system can undergo a transformation into an infinite dimensional space in which it evolves linearly in time.

To begin, consider an observable function $f$, per Eqn. (\ref{eqn:observable}). The Koopman operator, utilizing some transformation $T: S \rightarrow S$, is as follows:
    \begin{equation}
        \label{eqn:koopman-op}
        \mathcal{K}_T f(x) = f \circ T(x)
    \end{equation}

Since the Koopman operator is, mathematically, just a specific reformulation of the composition operator, its linearity follows simply \cite{arbabi2018introduction}:
    \begin{align}
    \begin{split}
        \label{eqn:koop-linearity}
        \mathcal{K}_T[g_1 + g_2](x) &= [g_1 + g_2] \circ T(x) \\
        &= g_1 \circ T(x) + g_2 \circ T(x) \\
        &= \mathcal{K}_T g_1(x) + \mathcal{K}_T g_2(x)
    \end{split}
    \end{align}

To apply the operator to our dynamical system, we use the system dynamics $\varphi$ as the transformation $T$. Now, instead of the system evolving nonlinearly in the state space from $z_t$ to $z_{t+1}$ (as in Eqn. (\ref{eqn:dyn-sys-map})), it evolves linearly in the space of the observables (with a slight abuse of notation in the last term, to make the evolution of the observable clear):
    \begin{equation}
        \label{eqn:koopman-form-dynamics}
        y_{t+1} = f(z_{t+1}) = \mathcal{K}_\varphi f(z_t) = \mathcal{K}_\varphi y_t
    \end{equation}
Complicating matters is the fact that since the Koopman operator evolves the system an infinite dimensional space, but this can be overcome by the ``somewhat naive but useful" \cite{arbabi2018introduction} ideation of the operator as a left-multiplication by infinite dimensional matrix. To make our problem computable, we can then make the assumption (common in data-driven approaches to the Koopman operator \cite{azencot2020forecasting}) that there exists a mapping $\chi$ that encodes the majority of the operator's action into a finite dimensional matrix $C$:
    \begin{equation}
        \label{eqn:C}
        C = \chi \circ \mathcal{K}_\varphi \circ \chi^{-1}
    \end{equation}
and conversely,
    \begin{equation}
        \label{eqn:C-chi}
        \chi^{-1} \circ C \circ \chi = \mathcal{K}_\varphi
    \end{equation}
    
Finally, we can notice that applying the Koopman operator repeatedly simplifies to an expression for predicting any $k \in \mathbb{N}$ number of steps into the future:
    \begin{align}
    \begin{split}
        \label{eqn:multistep-koopman}
            \mathcal{K}_\varphi \circ \mathcal{K}_\varphi \circ ...
        &= \chi^{-1} \circ C \circ \chi \circ \chi^{-1} \circ C \circ ... \\
        &= \chi^{-1} \circ C^k \circ \chi
    \end{split}
    \end{align}

Therefore we arrive at our formulation of the problem that we will utilize in building our network architectures:
    \begin{equation}
    \label{eqn:final-fwd-dynamics}
        y_{t+k} = \chi^{-1} \circ C^k \circ \chi (y_t)
    \end{equation}

If we properly construct networks in accordance with Eqn. (\ref{eqn:final-fwd-dynamics}), we know that those networks will function as good approximations of the Koopman operator. Thus we have applied physics-informed constraints to the network, though these constraints are informed by the physics of dynamical systems in general, not the specific physical laws governing a particular system\footnote{This is not a bad thing: This generality is precisely what makes Koopman networks so useful, \par because they can be applied to a multitude of different dynamical systems problems.}. The network is in turn more interpretable; we could, for example, analyze the encoder/decoder structure to better understand how the latent space relates to the space of the observables; or, we could calculate the eigenvalues of $C$ to gain insight into the system's stability.

\subsubsection{Convolutions}
    \begin{wrapfigure}{r}{0cm}
        \includegraphics[scale=0.4]{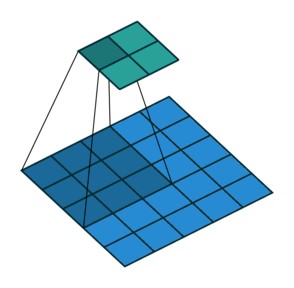}
        \caption{Example \\
        Convolution \protect\cite{sharfat2018convolutions}}
        \label{fig:convolution}
    \end{wrapfigure}
Convolutions are a common operation in machine learning on spatial data. To provide a basic intuitive description, convolution works by learning a kernel that condenses multiple datapoints into one; in Fig. \ref{fig:convolution}, the shaded 3x3 in the blue input layer is operated on by the kernel, producing the 1x1 shaded value in the green output layer. The rest of the output is generated in a similar way, by shifting the kernel along the dimensions of the input.

This operation can be viewed as a simple but effective incorporation of prior knowledge into machine learning models \cite{vonreuden2019knowledge}. Unlike fully-connected layers---where each node takes input from every node in the previous layer---convolutions enforce \textit{locality}, by taking input from spatially close nodes in the previous layer, and \textit{translation invariance}, by applying the same operation at all spatial locations in the data.

\subsection{Related Work}
\label{subsec:related-work}
There have been various approaches to SST forecasting with physics-agnostic machine learning models. Generic feed-forward neural network (NN) approaches to SST forecasting have been developed for at least two decades \cite{tangang1997forecasting}. With the the advent of deep learning techniques and hardware this decade \cite{lecun2015deeplearning}, deeper and more complex architectures have been used. In recent years, Long-Short Term Memory \cite{Xiao2019lstm}\cite{yang2018lstm}\cite{zhang2017lstm} and Autoencoder \cite{erichson2019shallow} architectures have been popular. Another interesting approach has been to learn to model the error of numerical models with a neural network \cite{patil2016numericalnn}.

Application of NNs to physics and earth science problems generally has a long history \cite{MILANO2002turbulent}\cite{baymani2014navier}\cite{reichstein2019earthsystem}, as an abundance of data and difficulty in accurately modeling certain systems has made the approach appealing. An expressly physics-informed approach to machine learning, however, was pioneered by Raissi et al. \cite{raissi2017physics}\cite{raissi2017physics2}, who utilized NNs to model partial differential equations describing physical systems. See Willard et al. \cite{willard2020integrating} for a thorough survey of the physics-informed ML approach.

Recently, the application of physics-informed ML to SST forecasting has received some attention. de Bezenec et al. \cite{debezenac2019deep} utilize physical knowledge of advection-diffusion to predict sea-surface temperature. Erichson et al. \cite{erichson2019lyapunov} incorporate constraints inspired by Lyapunov stability in their forecasting of various fluid flows, including sea-surface temperature.

Although the Koopman operator was developed in 1931 \cite{koopman1931hamiltonian}, it has recently begun to receive attention for its use in analyzing dynamical systems \cite{budisic2012applied}. This has spawned many machine-learning approaches to Koopman operator theory \cite{takeishi2017learning}\cite{otto2019linearly}. Our research here is inspired most immediately by Azencot et al.'s  \cite{azencot2020forecasting} recent work on ``consistent Koopman" networks, in which they use 180 day SST prediction as one of many tests for their novel method.

\section{Data and Method}
All code is implemented in Keras (v2.4) with a Tensorflow (v2.2) backend. Code is available at \url{https://github.com/JRice15/physics-informed-autoencoders}.

\subsection{Dataset}
\label{subsec:data}
We use the same dataset and configuration as Azencot et al. in their SST forecasting experiments. The dataset is the NOAA Optimal-Interpolation (OI) SST High-Resolution dataset \cite{reynolds2007oisst}, which measures daily snapshots of sea-surface temperature at $0.25^{\circ}$ spatial resolution globally, by interpolating satellite, buoy, and ship measurements to minimize bias. 

\begin{figure}[H]
        \centering
        \includegraphics[scale=0.55]{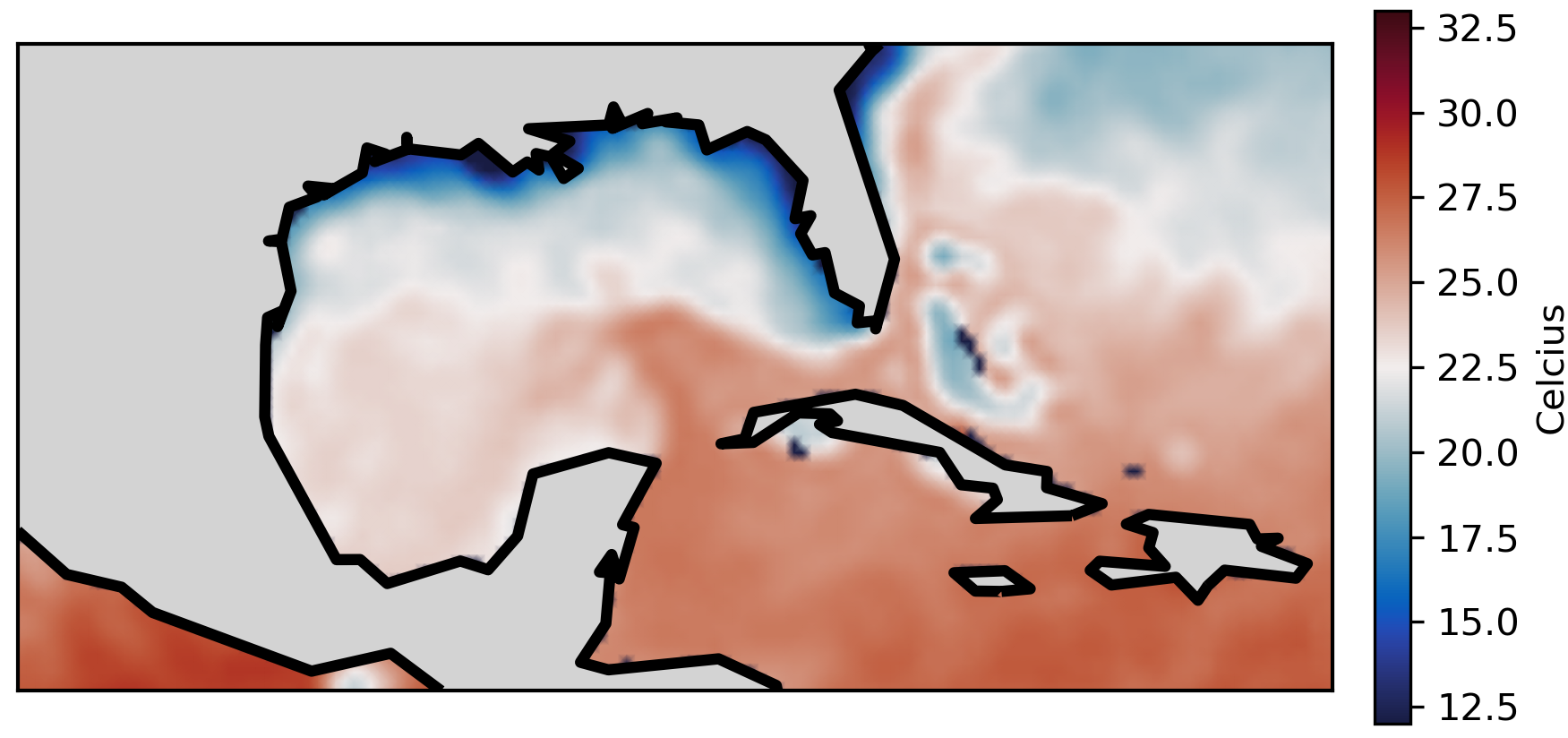}
        \caption{Example SST snapshot, day 140 in the training dataset}
        \label{fig:sst-input-140}
    \end{figure}

We focus on a 70x150 region of this dataset consisting of the Gulf of Mexico and a portion of the Caribbean (see Fig. \ref{fig:sst-input-140} for an example snapshot of this region). Though the thermal measurements are not actually photographic images, we may refer to them as ``images" and individual locations as ``pixels," as a useful shorthand.

We use three consecutive years (1095 days) as a training set, and the next year as a validation set during training. At test time, for longer term predictions, we use the three years following the training set as the test set (the slightly atypical overlap in validation and testing sets is to keep consistent with Azencot et al.'s experimental setup). The only difference from Azencot et al.'s setup is that used 30 days of testing data, while we use three years.

\subsubsection{Pre- and Post- Processing}
\label{subsubsec:pre-post-processing}
The measurements on land are set as zero, which cannot be confused as a temperature measurement as that sea surface temperature never occurs in the region. The data is then scaled to be approximately between -1 and 1. At test time, all locations where a zero (land) measurement appears are masked away, and do not contribute to error measurements. This is also done during training for convolutional networks, so that it does not contribute to the loss. We don't bother masking during training for fully-connected networks, as we find that it is trivial for them to learn the mask in the output layer (for example, by learning the weight to be zero and bias to the target mask value), however we still mask during testing so that averages are all computed identically.

\subsection{Physics Informed Autoencoders}
For more detailed descriptions of the following network architectures, see Appendix \ref{apndx:network-arch}. For training details, such as hardware used and learning rate schedules, see Appendix \ref{apndx:training}.

\subsubsection{Autoencoders: Introduction and Motivations}
To create a neural networks for SST prediction, we utilize an autoencoder (AE) framework. Autoencoders are a class of neural network that compresses data to a much smaller form without losing much information, by learning an \textit{encoder} (and its inverse, a \textit{decoder}). Looking at our formulation of the problem in Eqn. (\ref{eqn:final-fwd-dynamics}), one will be positively astounded to see that it naturally suggests such an autoencoder architecture, with $\chi$ and its inverse acting as an encoder and decoder respectively, for each of which we will substitute a clearer notation: $\chi_e$ and $\chi_d$. Furthermore, recent experiments (see Related Work, \ref{subsec:related-work}) suggest that sea-surface temperature evolution can be explained largely by low-dimensional dynamics, further pointing to an autoencoder approach.

Each network will thus consist of an encoder, $\chi_e$, a dynamics matrix than can be applied multiple times, $C$, and a decoder that maps back to the space of the observables, $\chi_d$:
\begin{equation}
    \label{eqn:koopman-ae}
    \hat{y}_{t+k} = \chi_d \circ C^k \circ \chi_e (y_t)
\end{equation}
We will use $k=14$ (days) for all of our experiments.

We develop and compare four implementations of Koopman autoencoders (KAEs) in sections \ref{subsubsec:simple-kae}--\ref{subsubsec:convcons-kae}, the relationships between which may be helpful to visualize:
\begin{table}[H]
    \begin{tabular}{|c|cc|}
        \hline
        & No Consistency & Consistency \\
        \hline
        Fully-Connected & Simple KAE & Cons KAE \\
        Convolutional & Conv KAE & ConvCons KAE \\
        \hline
    \end{tabular}
    \caption{Koopman Network Relationships}
    \label{tbl:koop-network-relationships}
\end{table}

\subsubsection{Learning the Koopman Operator}
In order for a network to learn a proper approximation of our Koopman-theoretic formulation in Eqn. (\ref{eqn:final-fwd-dynamics}), we first must instruct the network to generate accurate predictions given input $y_t$. It can do this by learning to minimize the difference between the ground truths $y_{t+1}$ and its predictions $\hat{y}_{t+1}$:
\begin{equation}
    \label{eqn:loss:one-step}
    \| \hat{y}_{t+1} - y_{t+1} \|_{\MSE}
\end{equation}
where $\| \cdot \|_{\MSE}$ denotes the mean-squared error (MSE), which squares its input element-wise and then averages over all dimensions and examples. However, since our goal is long-term prediction, we minimize the error over $k$ steps into the future:
\begin{equation}
    \label{eqn:loss:multi-step}
    \mathcal{L}_{pred} = \frac{1}{k} \sum_{n=1}^{k} \| \hat{y}_{t+n} - y_{t+n} \|_{\MSE}
\end{equation}

In addition to producing accurate predictions, we must also ensure that the network's elements are properly corresponding to their role in our Koopman-theoretic framework. For $C$, this is achieved by making it any linear operation, in our case matrix multiplication. To ensure that $\chi_e$ and $\chi_d$ serve only to encode and decode, and not learn any dynamics -- and conversely, that $C$ does not contribute to the encoding or decoding -- we implement a second loss term:
\begin{equation}
    \label{eqn:loss:identity}
    \mathcal{L}_{id} = \| \chi_d \circ \chi_e (y_{t}) - y_{t} \|_{\MSE}
\end{equation}
This enforces the relation that $\chi_d \circ \chi_e \approx I$, the identity.

The loss of the network is then the sum of the terms derived in Equations (\ref{eqn:loss:multi-step}) and (\ref{eqn:loss:identity}), weighted by respective hyperparameters $\lambda$:
\begin{equation}
    \label{eqn:loss:simple-full}
    \mathcal{L} = \lambda_{id} \mathcal{L}_{id} + \lambda_{pred}         
        \mathcal{L}_{pred}
\end{equation}
This loss (sometimes with a few additional terms) will be used by all four networks below.

\subsubsection{Simple KAE}
\label{subsubsec:simple-kae}
The Simple Koopman Autoencoder (SimpleKAE) is an autoencoder with three encoder layers, a dynamics operation, and three decoder layers. Each layer of the encoder or decoder is a standard fully-connected layer, consisting of a matrix multiplication of the weight and the input and addition of a bias term, followed by a hyperbolic tangent activation:
\[x_{out} = \tanh(W_i x + b_i)\]

\begin{figure*}[t]
    \includegraphics[scale=0.2]{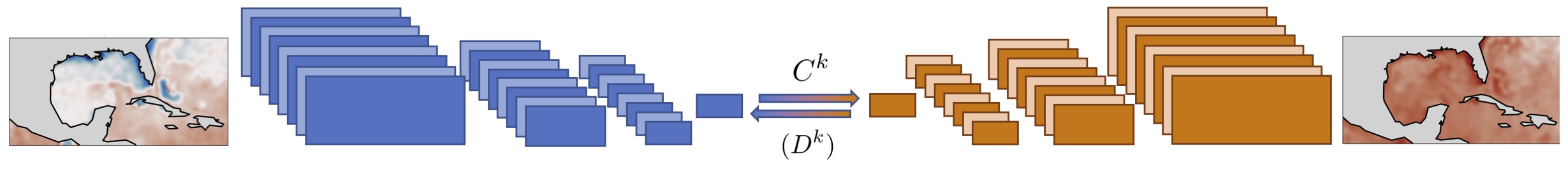}
    \centering
    \caption{Convolutional Network Architecture. The encoder translates a 70x150 snapshot into an encoded 1x24 representation in the latent space, which is transformed forward in time by the dynamics $C$ (and optionally backward by $D$, if it is a consistent Koopman network), and then decoded back to a 70x150 prediction.}
    \label{fig:conv-network}
\end{figure*}

\subsubsection{Consistent KAE}
The Consistent Koopman Autoencoder (ConsKAE), recently introduced by Azencot et al. \cite{azencot2020forecasting}, extends the SimpleKAE's structure based on the assumption that in the latent space, dynamics evolve linearly both forward and backward in time. This assumption is not necessarily true in all dynamical systems; entropic systems may evolve deterministically forward in time, yet it may be impossible to reconstruct past states from present ones. Testing consistent Koopman networks will help reveal to what extent this assumption applies to this SST domain.

For this network, we assume there exists a backward dynamics Koopman operator $\mathcal{K}_\psi$ in addition to the forward dynamics of $\mathcal{K}_\varphi$, such that
\begin{equation}
    \label{eqn:bwd-koopman}
    f(z_{t-1}) = \mathcal{K}_\psi f(z_t)
\end{equation}
$\forall t \in \mathbb{Z}$, where $\psi$ (the inverse of $\varphi$) maps $z_t$ to $z_{t-1}$. While we could technically use $C^{-1}$ to predict backward in our machine-learning approach, Azencot et al. find that the network is better able to learn backward dynamics via a seperate operation, $D$:
\begin{equation}
    \label{eqn:bwd-dynamics}
    \hat{y}_{t-k} = \chi_d \circ D^k \circ \chi_e (y_t)
\end{equation}
The error of backward predictions is computed analogously to Eqn. (\ref{eqn:loss:multi-step}) for another loss term, $\mathcal{L}_{bwd}$. Additionally, in order to enforce that the backward and forward dynamics are \textit{consistent}---meaning that they are approximate inverses---we use the following loss term:
\begin{align}\begin{split}
    \label{eqn:loss:bwd}
    \mathcal{L}_{cons} = \sum_{n=1}^\kappa & \left[ \frac{1}{2n} \|D_{n*} C_{*n} - I_n\|_{F}^2 \right. \\
    + & \left.\frac{1}{2n} \|C_{n*} D_{*n} - I_n\|_{F}^2 \right]
\end{split}\end{align}
where $D_{n*}$ denotes the upper $n$ rows, and $D_{*n}$ the leftmost $n$ columns, of D. The $n \times n$ identity matrix is represented by $I_n$, and $\|\cdot\|_{F}$ is the Frobenius norm. For the derivation of this term, see  section 4 of Azencot et al. \cite{azencot2020forecasting}.

The ConsKAE thus uses the following loss, with the two new loss terms weighted by respective $\lambda$ hyperparameters:
\begin{equation}
    \label{eqn:loss:full-consistent}
    \mathcal{L} = \lambda_{id} \mathcal{L}_{id} + \lambda_{pred}         
        \mathcal{L}_{pred} + \lambda_{bwd} \mathcal{L}_{bwd} + \lambda_{cons}
        \mathcal{L}_{cons}
\end{equation}

Azencot et al. \cite{azencot2020forecasting} tested their model on this same SST domain, and was our inspiration for doing the same. We will compare their results with ours, though we utilized what we believe to be a more robust and informative testing scheme.

\subsubsection{Convolutional KAE}
The Convolutional Koopman Autoencoder (ConvKAE) uses the same dynamics as the SimpleKAE, but uses a convolutional encoder/decoder instead of fully-connected layers. The encoder consists of four layers, the first three of which consist of 16-filter convolutions and hyperbolic tangent activations followed by max-pooling, while the last is a 1-filter convolution alone. The decoder is the reverse, with transposed convolutions and upsampling replacing convolutions and max-pooling, respectively. See Fig. (\ref{fig:conv-network}) for a graphic representation.

\subsubsection{Convolutional-Consistent KAE}
\label{subsubsec:convcons-kae}
A combination of the previous two implementations, the Convolutional Consistent KAE (ConvConsKAE) uses both the convolutional encoding/decoding scheme of the Convolutional KAE, and paired forward and backward dynamics of the Consistent KAE.

\subsection{Baselines}
\subsubsection{Persistence}
Persistence is what it sounds like: a model that passes its input to its output, unchanged. Comparison to this baseline reveals the extent to which a network learns the actual dynamics of the system, and not simply a good encoding-decoding scheme.
\subsubsection{Learned Average}
The Learned Average model (LrndAvg) learns and outputs a temporally constant spatial distribution that is the same shape as its input. This distribution is learned by minimizing the mean-squared error between its inputs and its output, so that over the course of training a value akin to the weighted average at each spatial location is learned. Comparison to LrndAvg reveals whether has devolved to guessing the average and is not taking its input into account---or, perhaps even worse, whether it is making a legitimate effort that is still no better than guessing the average. 

\subsection{Testing Methods and Metrics}
We trained each model on 12 different seed values, meaning 12 trials with unique network initializations. To test the models for long-term prediction, we record each seed's average performance on 180 day prediction over three years of test data (see section \ref{subsec:data} for description of this dataset).

In testing, we evaluate the following metrics, in terms of a prediction $\hat{y}_t$ and target $y_t$.
\begin{itemize}
\item Relative prediction error (RelPred) measures the error scaled by the intensity of the target:
    \begin{equation}
        \label{eqn:relpred}
        E_{relpred} = \frac{\| \hat{y}_t - y_t \|_F}{\| y_t \|_F}
    \end{equation}
This is used so that the error of hot and cold months can be compared on more equal footing.
\item Celcius Mean-Absolute Error (MAE) is simply the average absolute difference between $\hat{y}_t$ and $y_t$, in units that are universally interpretable (unlike relative prediction error). We will prefer these units because of their universal meaning.
\end{itemize}
We also measured mean-squared error and mean-absolute error, though they will not appear here as we find they did not seem to provide any insights not already represented in the two metrics above.

We calculate not only multiple metrics but multiple aggregations of those metrics among the 12 trials of each model: average, median, min, max, and standard deviation.

\section{Experimental Results}

\subsection{Baselines}
\begin{figure}[H]
    \includegraphics[scale=0.60]{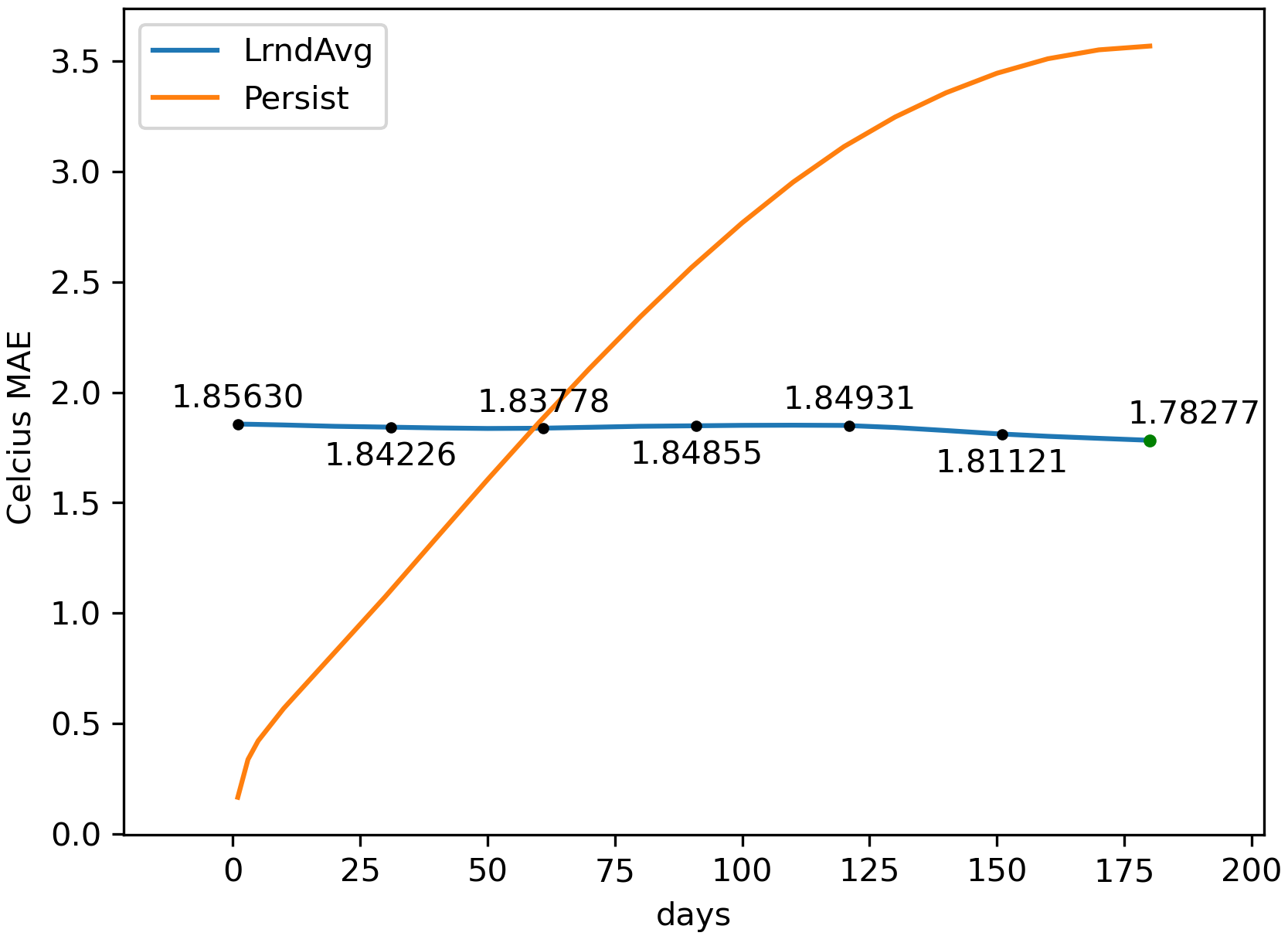}
    \caption{Celcius MAE of Baselines}
    \label{fig:baselines-celcius}
\end{figure}

To begin, we evaluate the baselines' performances in Fig. \ref{fig:baselines-celcius}. LrndAvg stays relatively flat (as expected), with the error hovering around 1.8 degrees. The Persistence model starts off well, but rapidly diverges to almost twice the error of the LrndAvg. While the LrndAvg model represents a good target to beat, Persistence is more of an upper-bound on how reasonably \textit{poor} a model can perform.

\subsection{Koopman Autoencoders}

\subsubsection{Initial Observations}
\begin{figure}[H]
    \includegraphics[scale=0.60]{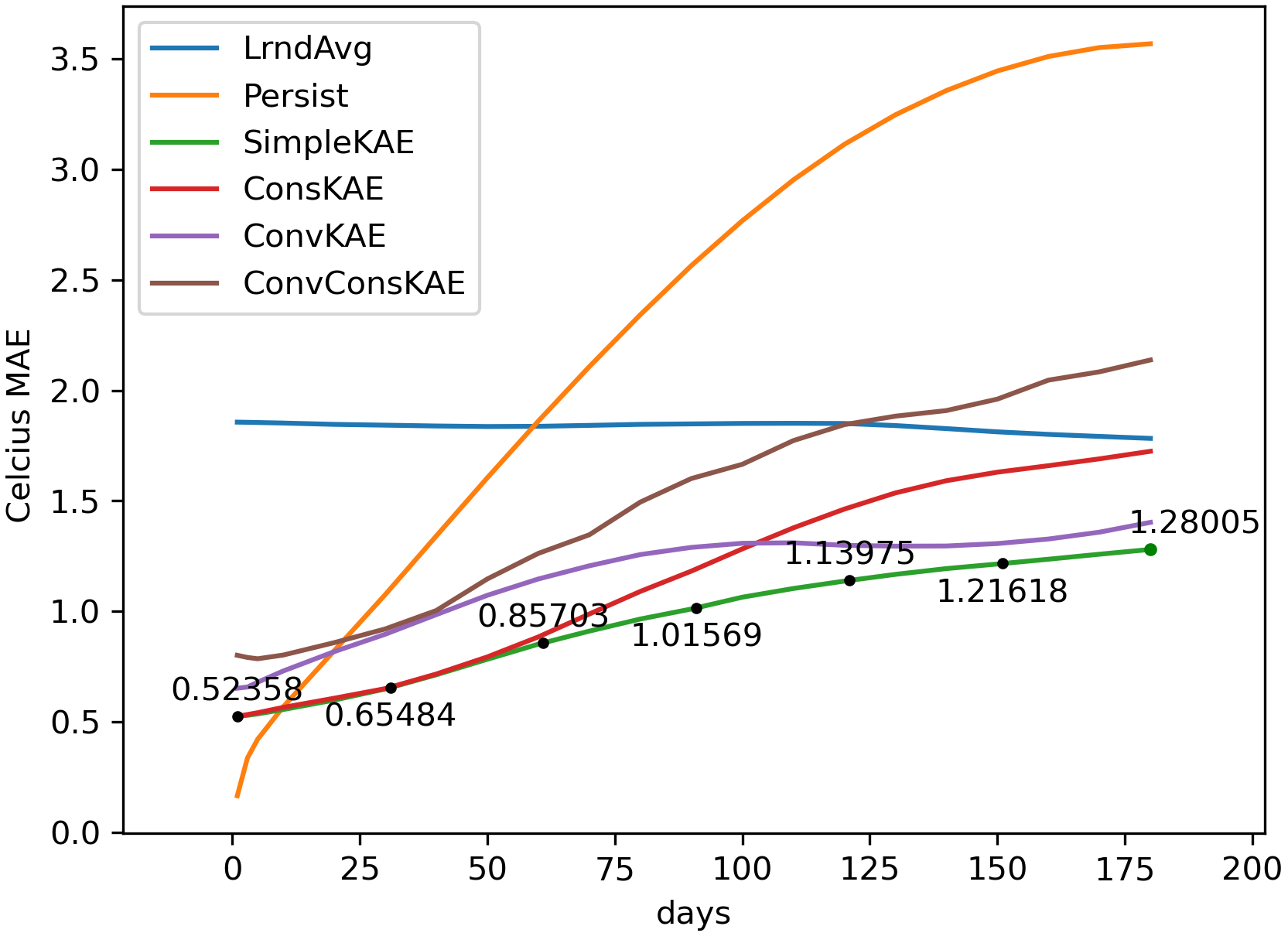}
    \caption{Average Celcius MAE of Koopman Models}
    \label{fig:avg-models-celcius}
\end{figure}

\begin{figure}[H]
    \includegraphics[scale=0.60]{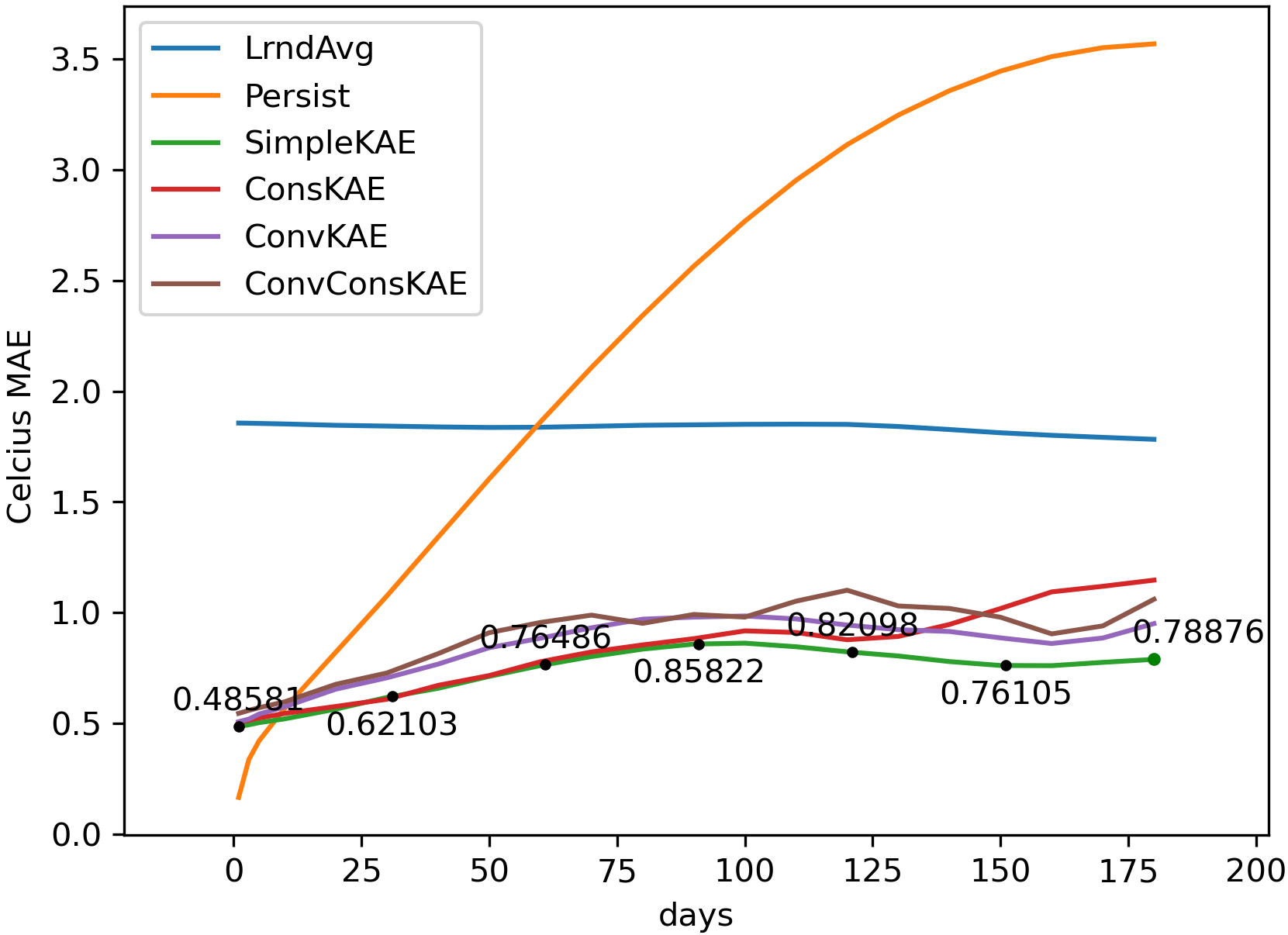}
    \caption{Minimum Celcius MAE of Koopman Models}
    \label{fig:min-models-celcius}
\end{figure}

In our experiments, the SimpleKAE appears to have performed the best, followed by the ConvKAE, ConsKAE, and lastly ConvConsKAE. This order is present in the average (Fig. \ref{fig:avg-models-celcius}) and median aggregations of model seeds. It is the same according to the minimum aggregation (Fig. \ref{fig:min-models-celcius}), except that ConvConsKAE performs better ConsKAE. 

\begin{figure*}[t]
    \includegraphics[scale=0.49]{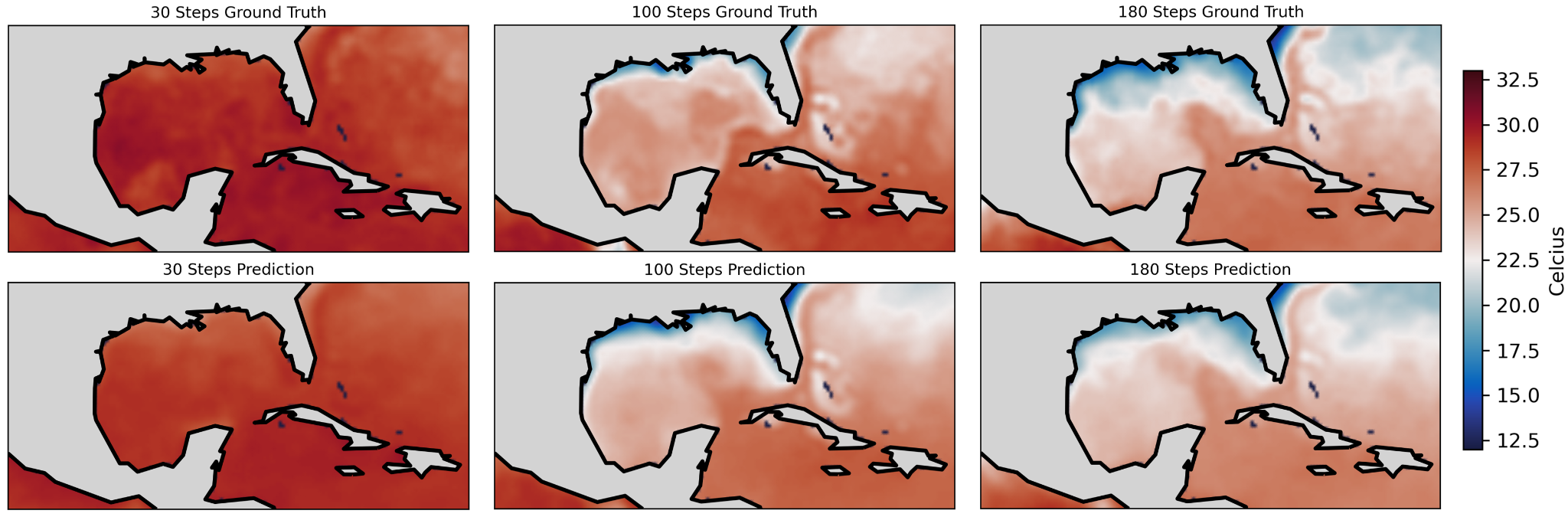}
    \caption{Selected day 30, 100, and 180 ground truths (top) and best SimpleKAE seed predictions (bottom). Though the predictions are lacking the fine-grain detail of the ground truth---which is understandable given that the predictions are decoded from a vector of length 12---the predictions follow the ground truth remarkably closely on the large scale. Note the similarity to ground truth at day 180, including a fairly accurate prediction of both the Loop Current and the Gulf Stream.}
    \label{fig:best-preds}
\end{figure*}

Fig. \ref{fig:best-preds} shows visual results of the best performing SimpleKAE seed, for predictions of 30, 100, and 180 days into the future. It demonstrates the model's ability to accurately predict seasonal changes, and the existence of key features including the Loop Current and Gulf Stream. However it does lack finer detail, which is to be expected given how far into the future it is trying to predict.

The median and maximum errors of each model finish in identical order to the average. It should be noted that the median is noticeably lower than the average, suggesting high outliers. Using the relative prediction error metric yields a very similar picture, enough so that it need not be included.

\subsubsection{Analysis of Variance}
\begin{figure}[H]
    \includegraphics[scale=0.6]{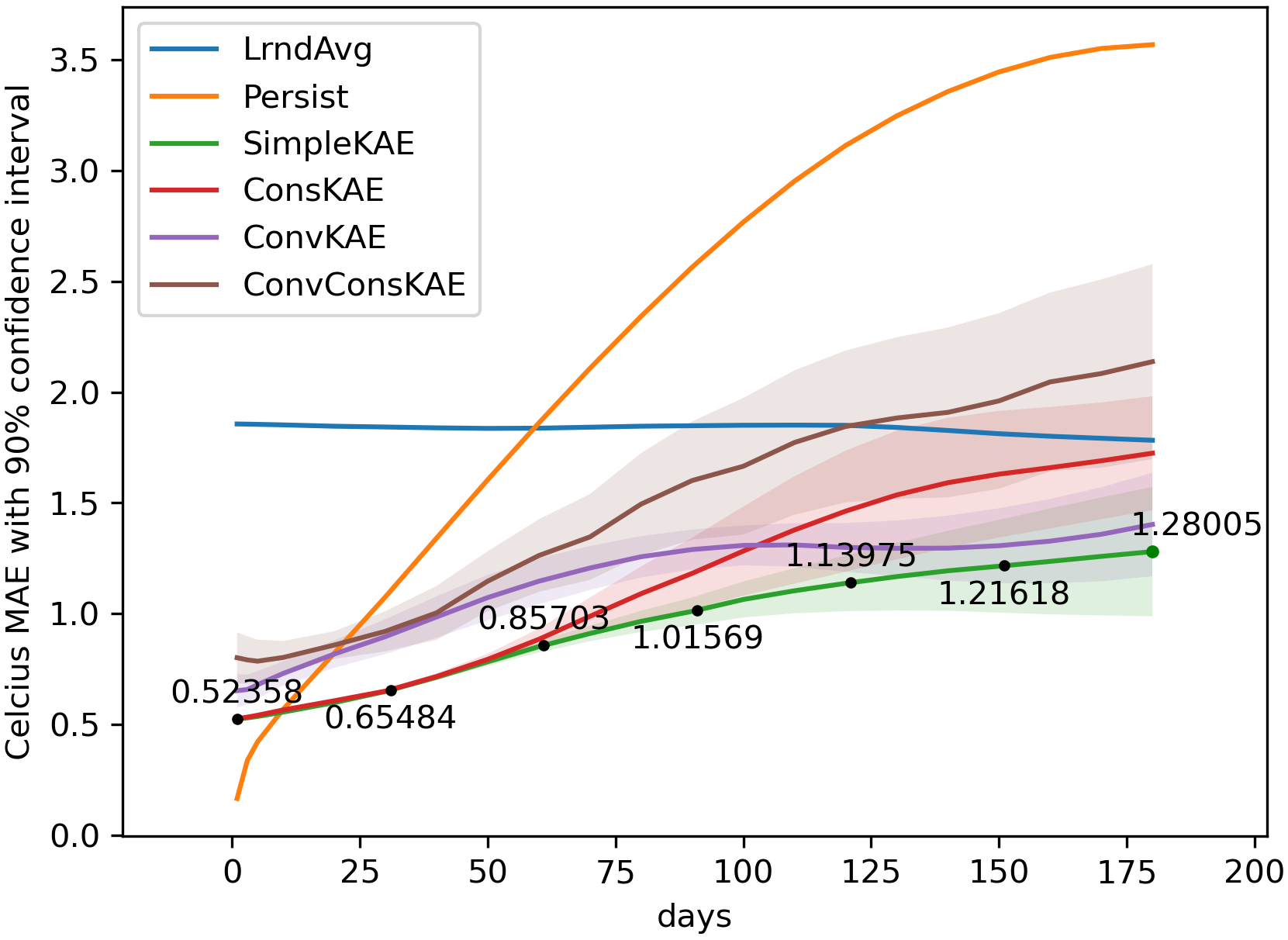}
    \caption{Average Celcius MAE with 90\% Confidence}
    \label{fig:avg-models-celcius-w-confidence}
\end{figure}

Complicating the above results, we must note that there is a relatively large variance within each model's 12 seeds. Take for example the spread of the best performing model's seeds, in Fig. \ref{fig:simplekae-seeds}, which confirms the existence of high outliers:
\begin{figure}[H]
    \includegraphics[scale=0.60]{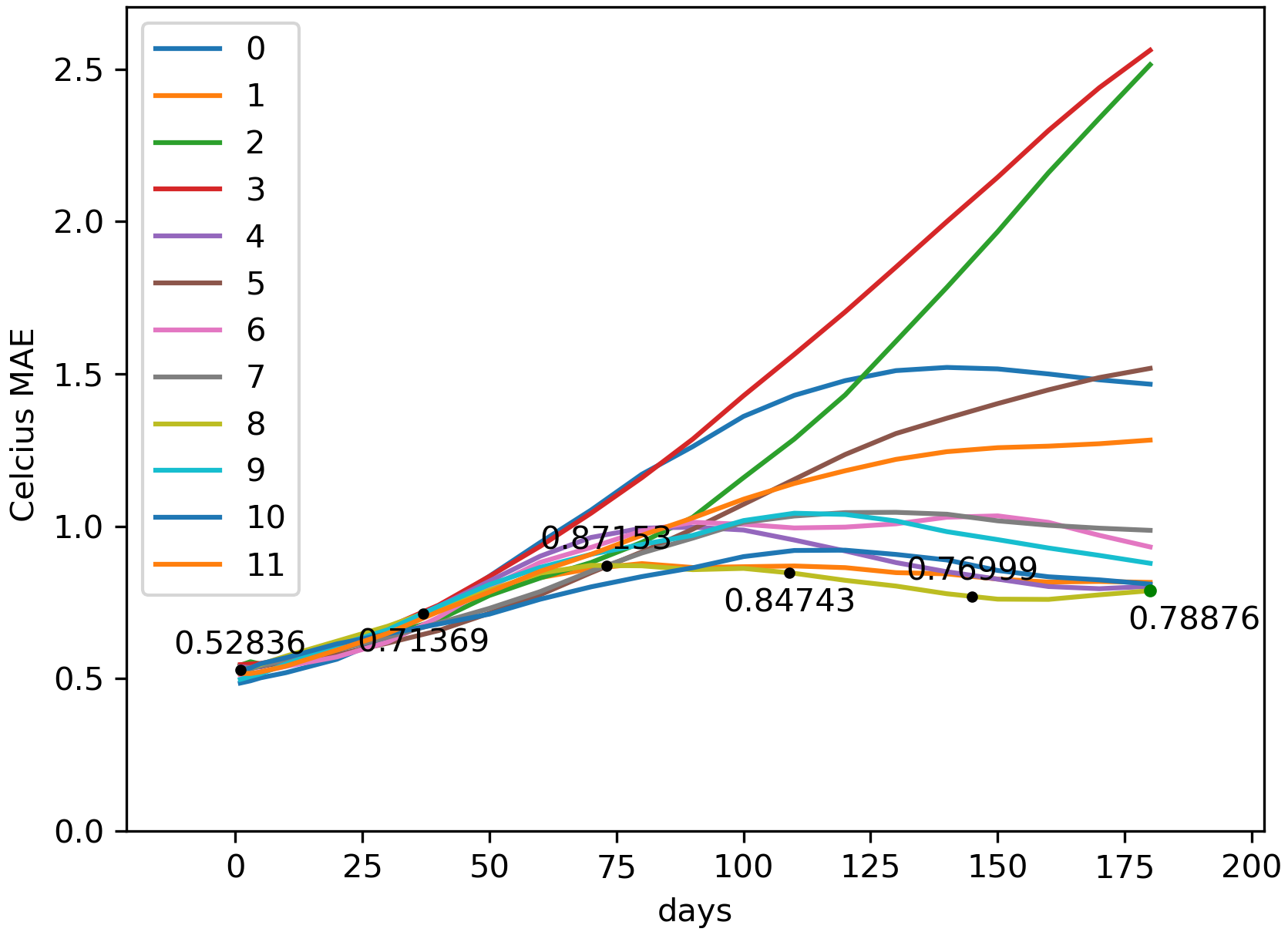}
    \caption{SimpleKAE Seeds}
    \label{fig:simplekae-seeds}
\end{figure}
There is great variation around their mean of $1.28^{\circ}$ Celcius, most notably on the high end with some seeds finishing at double the mean. Interestingly, it is nearly impossible to tell from their prediction performance during the first 25 or so days whether a given seed will perform well or poorly at 180 days. Similarly, training performance is not very indicative of how well a seed will perform at long-term prediction; all models achieved very similar training losses (which takes into account only 14 prediction days). This suggests that during training, the model is able to find minima in the optimization space for each seed that perform similarly in the short term, but that some of those minima are much more or less stable than others for long-term forecasting. 
Fig. \ref{fig:avg-models-celcius-w-confidence} shows all models' variances, with 90\% confidence bounds calculated using the standard deviation of the seeds from their mean.
Though it is a bit messy, the key conclusion to draw from Fig. \ref{fig:avg-models-celcius-w-confidence} is that many confidence intervals overlap with other means, suggesting that we may not be able to say we are certain they are significantly different. To this end, we apply Welch's t-test \cite{welch1947t-test} to determine how confident we are that the day 180 means are in fact different:
\begin{table}[H]
    \centering
    \begin{tabular}{|c|c|c|c|c|}
        \hline
    	& LrndAvg & Simple & Cons & Conv	\\
    	\hline
        LrndAvg	& - & 98.375 & \textbf{28.378} & 97.853	\\
        Simple & 98.375 & - & 92.610 & \textbf{40.467} \\
        Cons & \textbf{28.378} & 92.610 & - & \textbf{85.780} \\
        Conv & 97.853 & \textbf{40.467} & \textbf{85.780} & -	\\
        ConvCons & \textbf{78.852} & 98.497 & \textbf{80.076} & 97.320 \\
        \hline
    \end{tabular}
    \caption{Welch t-Test Confidences (two-tailed; values in percent) for the difference between means. We consider $\ge 90\%$ to be significant. Methods with non-significant differences are bolded.}
    \label{tbl:welch-t-test}
\end{table}
Table \ref{tbl:welch-t-test} shows the confidences that the means of any two methods are in fact significantly different. We chose a confidence of 90\% because of our small sample sizes, though a 95\% confidence would not have made drastic changes (only the SimpleKAE v. ConsKAE significance).

From this data, we cannot conclude that the SimpleKAE performs any better than the ConvKAE. In fact, for all Koopman methods, we can see that the differences between one method and the next best method (according to their day 180 average MAE; see Fig. \ref{fig:avg-models-celcius}) are not significant. However, analysis in terms of the models' relationships per Table \ref{tbl:koop-network-relationships} reveals that there are statistically significant differences between consistent Koopman methods and their non-consistent counterparts (though there are not significant differences between non-convolutional and convolutional methods), at least at 90\% confidence. Given that using 95\% confidence bound would change this relation, we cannot be very confident that our results do simply reflect a bad batch of seeds for the consistent methods.

\section{Discussion}

Our results consistently validate a Koopman-theoretic approach to long-term SST forecasting, a challenging dynamical systems problem. The best initializations of all Koopman methods significantly outperformed both the Persistence and LrndAvg baselines. Qualitatively, our best performing models seem to capture the general trends of seasonality, as well as some fine grain features. Although it may seem strange that we are able to achieve such results from encoding a single thermal image to a vector of length 12 or 24 and projecting 180 days into the future, Azencot et al. note (and we experimentally validate) that larger-scale SST dynamics are empirically low-dimensional, and can be modelled with the Koopman operator.

Our results suggest that the consistent Koopman methods do not tend to perform as well as their non-consistent baselines in this SST domain. This is contradictory to the results of Azencot et al. \cite{azencot2020forecasting}, who reported the following data:

\begin{table}[H]
\centering
\begin{tabular}{|c|ccc|}
    \hline
    Model & \multicolumn{3}{c|}{RelPred Error} \\
    & Min & Max & Avg. \\
    \hline
    SimpleKAE & 0.315 & 1.062 & 0.701 \\
    ConsKAE & 0.269 & 1.028 & 0.632 \\
    \hline
\end{tabular}
\caption{Day 180 Relative Prediction Error on SST Dataset, with 18 seeds \protect\cite{azencot2020forecasting}}
\label{fig:azencot-relpred}
\end{table}

They report an improvement in relative prediction error of 0.069, but with a variation between min and max of 0.759. Without further information, it is difficult to know what significance to assign these results. We find the following, using the same relative prediction error metric:

\begin{table}[H]
\centering
\begin{tabular}{|c|ccc|}
    \hline
    Model & \multicolumn{3}{c|}{RelPred Error} \\
    & Min & Max & Avg. \\
    \hline
    SimpleKAE & 0.292 & 0.863 & 0.451 \\
    ConsKAE & 0.412 & 1.151 & 0.583 \\
    \hline
\end{tabular}
\caption{Our Day 180 Relative Prediction Error, with 12 seeds}
\label{fig:our-relpred}
\end{table}
We find approximately double the difference between means but in the opposite direction, though we do find a similar min-max difference. We attribute the differences to both our more robust testing scheme; while Azencot et al. test on the 30 days following the training set, we test on the three years following. However, they tested 18 seeds to our 12, suggesting our results may be more susceptible to a bad batch of seeds. There are also slight differences in the number of epochs trained and learning rate schedules of the models, though we do not expect these differences to have a great effect.

On other datasets where it is known that consistency applies---such as computer-generated data of ideal fluid flow past a cylinder, where we know that past states can be predicted from the present just as easily as future states---Azencot et al. demonstrated significant improvements over the SimpleKAE. Their results are much less conclusive in the more challenging real-world domain of SST forecasting, however. Given the unexpectedly large variation between seeds and our differing results, it is hard to draw conclusions as to whether the consistent method is better than the its simple counterpart. At the very least, it seems reasonable to say that Koopman consistency is not especially beneficial for forecasting in this domain---though possibly not especially harmful either.

Our experiments also suggest that the assumptions inherent in a convolutional encoder/decoder architecture are not especially beneficial. Although the \textit{locality} of the data seems intuitively to be correct, it is possible that assuming \textit{translation invariance} in the encoding scheme can make it more difficult to create encodings that extract the important information from all regions---e.g., the same filter may not be able to extract the most useful encoding from both the coast of Louisiana and the surroundings of Cuba.

Thus our primary conclusion is that informed machine-learning is only beneficial when the \textit{right} prior information and assumptions are used. Though our experiments have validated the effectiveness of a Koopman-theoretic method, the additional assumptions inherent in convolutional and consistent Koopman methods do not seem to be especially beneficial in the complex and challenging domain of long-term sea-surface temperature forecasting.

Opportunities for future work include the validation of our methods in other regions and with other datasets, and testing alternative model architectures and configurations. A particularly interesting avenue, given our goal of long-term prediction, could be construct the model to predict multiple days each step (e.g. predict in steps of 5 or 10 days, instead of 1). This could improve computation time, reduce the number of steps during which errors could get compounded, and possibly make simply learning the identity function a less appealing option for the model.

\section{Acknowledgements}
This work was supported in part by the U.S. Department of Energy, Office of Science, Office of Workforce Development for Teachers and Scientists (WDTS) under the Science Undergraduate Laboratory Internships Program (SULI).

Many thanks to N. Benjamin Erichson for his guidance in understanding his recent paper.

Thanks also to Pacific Northwest National Laboratory, and specifically to PNNL Research Computing for giving us access to computing resources to complete this project.

We would like acknowledge Google Cloud Computing for supporting this project with compute credits.

\section{Data Availability Statement}
The data that support the findings of this study are available from the corresponding author upon reasonable request. Additionally, all code is available at \url{https://github.com/JRice15/physics-informed-autoencoders}.

\bibliographystyle{plain}

\bibliography{references}

\end{multicols}

\appendix

\section{Network Architectures}
\label{apndx:network-arch}
\begin{multicols}{2}

Other than dynamics layers, all weights are initialized with the Glorot Normal (aka Xavier Normal) initializer \cite{pmlr-v9-glorot10a}, while all biases are initialized to zero. Forward dynamics weights are initialized by taking the singular value decomposition of a matrix generated via a random normal distribution, and then multiplying $U$ and $V^T$ to create the weight. Backward dynamics weights, when applicable, are initialized with the pseudo-inverse of the forward weights' transpose. Neither dynamics use a bias term.
\end{multicols}

\subsection{Fully-Connected Networks}
\begin{table}[H]
\centering
\begin{tabular}{|c|c|c|c|}
    \hline
    Component & Layer & Activation & (Output) Shape \\
    
    \hline
    Input & - & - & 10500 \\
    
    \hline
    Encoder & Fully-Connected 1 & tanh & 96 \\
     & Fully-Connected 2 & tanh & 96 \\
     & Fully-Connected 3 & - & 12 \\

    \hline 
    Dynamics & Fully-Connected & - & 12 \\

    \hline
    Decoder & Fully-Connected 1 & tanh & 96 \\
     & Fully-Connected 2 & tanh & 96 \\
     & Fully-Connected 3 & - & 10500 \\
    \hline
\end{tabular}
\caption{Fully-Connected Architecture. This architecture was developed by Azencot et al \cite{azencot2020forecasting}. They performed a parameter search when developing this model.}
\end{table}

\vspace{1em}

\subsection{Convolutional Networks}

\begin{table}[H]
\centering
\begin{tabular}{|c|c|c|c|c|c|}
    \hline
    Component & Layer & Kernel Size & Size Modification & Activation & (Output) Shape \\
    \hline
    Input & - & - & - & - & 70x150 \\
    \hline
    
    Encoder & Convolution 1 & 3x3 & 2x2 max-pooling & tanh & 34x74x16 \\
    & Convolution 2 & 3x3 & 2x2 max-pooling & tanh & 16x36x16 \\
    & Convolution 3 & 5x5 & 2x2 max-pooling & tanh & 6x16x16 \\
    & Convolution 4 & 5x5 & flattening & - & 24 \\
    \hline
    
    Dynamics & Fully-Connected & - & - & - & 24 \\
    \hline
    
    Decoder & Zero Padding & - & unflatten, 0x0x1x1 padding & - & 7x17 \\
    & Convolution 1 & 5x5 & - & tanh & 7x17x16 \\
    & Convolution 2 & 5x5 & 2x2 upsampling & tanh & 22x42x16 \\
    & Convolution 3 & 3x3 & 2x2 upsampling & tanh & 48x88x16 \\
    & Convolution 4 & 3x3 & 2x2 upsampling & - & 100x180 \\
    & Cropping \& Masking & - & crop to 70x150 & - & 70x150 \\
    \hline
\end{tabular}
\caption{Our Convolutional Architecture}
\end{table}

\begin{multicols}{2}
\vspace{1em}
At the beginning of the decoder, one row and column of zero padding is applied to the bottom and right of the 6x16 snapshot, and then at the end of the decoder the output is cropped to the target shape, to make the input and output shapes of the AE the same. All locations where land is present is masked away in both the network's output and in the target images, so that it does not affect the loss (see section \ref{subsubsec:pre-post-processing}). All upsampling uses the bilinear interpolation technique. Output shape is given in the form (height x width x channels).

Though we did not perform an exhaustive parameter search, we did explore many different avenues when developing this architecture. Here are some of the options we explored, and the reason why we did not use them when it was more complex than simply worse performance, as guidance for those seeking to improve upon our results:
\begin{itemize}
    \item Batch normalization: we suspect this failed because of the seasonality of SST. When the data is normalized, it can be more difficult to tell the difference between warmer and colder states that are equally distributed.
    \item ReLU activation: We suspect that ReLU performed poorly because it loses information by design, which we want to minimize in the AE framework. Additionally, because of its simplicity relative to the hyperbolic tangent, we would likely need a deeper network to get similar results. We found the ReLU often devolved toward the LrndAvg's output.
    \item Various numbers of convolution output channels. We use 16/16/16/1; with less than that, such as 6/6/6/1, the AEs output was not as smooth, and did not perform as well. With larger than that, such as 16/32/64/1, the model became unstable, and would often diverge suddenly during long-term prediction at test time.
    \item Convolutional dynamics: we explored this briefly, since convolution is also a linear operator, but it performed very poorly. 
    \item Average pooling and nearest-neighbor upsampling
    \item Kernel size ordering: we use the larger kernels for the more compressed sizes, to reduce the number of operations and thus increase the speed of the network.
\end{itemize}

\section{Training}
\label{apndx:training}
Nvidia Tesla K80 and P100-PCIE GPUs were used for training. 14 forward prediction steps (and 14 backward, for consistent KAEs) were used for training losses. All models were trained for 2000 epochs (with early stopping) with the Adam optimizer, 0.5 gradient norm clipping, and a batch size of 64. Learning rate was on a schedule, starting at 0.01 and reducing by a factor of 0.6 ($lr_{new} = 0.4 \times lr_{old}$) at epochs 50, 400, 800, 1200, and 1600. Early stopping came into effect if 420 epochs passed with no improvement in validation loss.
\end{multicols}

\end{document}